\documentstyle[multicol,aps]{revtex}
\begin{document}
\title{Avalanche dynamics and nonexponential relaxation}
\author{Alexei V\'{a}zquez$^1$, Oscar Sotolongo-Costa$^{1,2}$, Carlos Antoranz$^2$.}
\address{1.-Department of Theoretical Physics. Havana University.\\
Havana 10400, Cuba.\\
2.-\ Departamento de Fisica Fundamental, Fac. Ciencias,\\
UNED, Madrid 28080, Spain}
\date{\today}
\maketitle

\begin{abstract}
The theory of SOC, and related avalanche dynamics, is proposed as the origin
of the ubiquitous nonexponential relaxation observed in complex systems.
Introducing some scaling laws and relations we have obtained that the
normalized relaxation function follows an stretched exponential decay and
that the frequency spectrum follows an ''$1/f$'' noise. Moreover, in the MF
approach the relaxation is found to be exponential.
\end{abstract}


\begin{multicols}{2}

\section{Introduction}

Self-Organized criticality (SOC) was proposed as a general theory to
understand the ubiquitous ''$1/f$'' noise \cite{weissman} and self-similar
fractal  structures \cite{mandelbrot}, in dynamical systems with many degrees
 of freedom. The SOC idea has been illustrated by a large variety of computer
models, including sandpiles \cite{bak}, evolution \cite{evolution}, interface
depinning \cite{interface}, and more \cite{more}, in which a slow driving
field leads to a stationary state with avalanches with widely distributed
sizes.  However, most of the research has been devoted to the study of
 the stationary critical state.

Another puzzle seeking a physical explanation is the empirical observation
 that relaxation phenomena in nature is not exponential. Nonexponential
 relaxation has been observed in different phenomena like mechanical,
 dielectric and magnetic relaxation, and in a large variety of materials,
 including amorphous polymers, supercooled liquids, glasses, spin glasses,
 and amorphous semiconductors \cite{relaxation}. These are complex systems
 where relaxation phenomena offer a large variety of challenging problems.

Many theoretical models has been proposed to explain the ubiquitous
 nonexponential relaxation \cite{percolation,cooperative,constrained,trapping
}. The physical origin of the nonexponential decay is attributed to the
 existence of fractal clusters \cite{percolation}, cooperative dynamics
 \cite{cooperative}, hierarchical constraints \cite{constrained},
 or anomalous diffusion \cite{trapping}. All these models have
 their advantages and limitations and may be applied to some,
 but not all, experimental situations. 

In the present work we propose the SOC theory, and the related
 avalanche dynamics, as a physical mechanism leading to the nonexponential
 relaxation, often observed in nature. The present approach may be
 enclosed among those models which attribute the nonexponential behavior
 to the existence of fractal clusters, however the clusters (avalanches)
 in the SOC have an intrinsic dynamical nature, and correlations are present
 not only in space but also in time.

\section{Critical relaxation}

In this section we analyze the relaxation of a self-organizing system near
 the critical state. We first start with a mean-field approach to the problem
 and later we analyze some scaling laws for the different magnitudes
 characterizing the dynamics of the system.

\subsection{Mean field theory}

The first step towards a comprehensive understanding of SOC and related
 phenomena is provided by mean-field (MF) theories, which give insight into
 the fundamental physical mechanism of the problem. The first MF analysis
 to sandpile models is due to Tang and Bak \cite{tang}. They mapped a
 critical height (Abelian) sandpile model in a two state cellular automaton
 and took the driving rate and the average height as control parameters.
 Later Caldarelli, Maritan and Vendruscolo  generalized the MF approach of
 Tang and Bak in order to include stochastic rules\cite{caldarelli}. On the other hand,
 Vespignani and Zapperi \cite{vespignani} mapped the sandpile automata in a
 three state cellular automata on a $d$-dimensional lattice, and took the
 driving field $h$ and the dissipation $\epsilon$ rates as control parameters.
 In this way, they obtained that the system is critical in the double limit
 $h,\ \epsilon\rightarrow 0$ and $h/\epsilon\rightarrow 0$. The approach 
developed here is in the same line as that of Vespignani and Zapperi \cite{vespignani}.

We divide the states that each site can assume in stable, unstable and active, 
and denote their average densities by $\rho_s$, $\rho_u$ and $\rho_a$, respectively. 
Stable sites are those that cannot become active by addition of grains. Unstable sites
 are those that may become active by addition of grains. Active sites are
 relaxing and transfer energy to its nearest neighbors. Sites may receive energy either
 from the driving field or from its nearest active neighbors. The driving field is
 characterized by the driving rate $h$, the probability per unit time that a site will receive
 a grain of energy from the driving field. On the other hand, the probability per unit time that a
 site receives a grain from nearest active neighbors is given
 by $1-(1-\rho_a)^g\simeq g\rho_a$, where $g$ is the effective number of
 nearest neighbors and $\rho_a$ is small. Then, taking into account both contributions, 
the local driving field is given by 
\begin{equation}
h_{\text{loc}}=h+g\rho_a\ .
\label{eq:1}
\end{equation}
This local field gives the probability per unit time that a site receive a grain, either
 form the driving field or from nearest active neighbors.

Let $u$ be the fraction of stable sites that become unstable by addition of a 
grain, $p$ the fraction of unstable sites that become active by addition of a grain, 
and $q$ the fraction of active sites that become stable. With these definitions 
we write the following rate equations for the average densities
\begin{equation}
\frac{d}{d t} \rho_a = -[1-(1-q)ph_{\text{loc}}]\rho_a + ph_{\text{loc}}\rho_u\ ,
\label{eq:2}
\end{equation}
\begin{equation}
\frac{d}{d t} \rho_s = q\rho_a - uh_{\text{loc}}\rho_s\ ,
\label{eq:3}
\end{equation}
plus the normalization condition 
\begin{equation}
\rho_s+\rho_u+\rho_a=1\ .
\label{eq:4}
\end{equation}
The first term in the right hand side of eq. (\ref{eq:2}) characterizes the transition of active
 sites to stable or unstable sites. The correction $(1-q)ph_{\text{loc}}$ takes into
 account that the fraction $1-q$ of active sites that become unstable may remain active
 by addition of grains from the local field. The second term of the same equation
 characterizes the transition of unstable sites to active induced by the local driving field. 
On the other hand, the first term in the right hand side of eq. (\ref{eq:3}) gives the
 fraction of active sites that become stable, and the second one, the transition of stable
 sites to unstable ones induced by the local driving field.

After imposing stationarity ($\frac{d}{d t}\rho_a=0$,
 $\frac{d}{d t}\rho_s=0$), from eqs. (\ref{eq:1}-\ref{eq:4}) one obtains
\begin{equation}
upqg\rho_a^2 + \big[pq+u(1-pg+pqh)\big]\rho_a - uph = 0\ ,
\label{eq:5a}
\end{equation}
\begin{equation}
\rho_s = \frac{q}{u}\frac{\rho_a}{h+g\rho_a}\ ,
\label{eq:5b}
\end{equation}
\begin{equation}
\rho_u = 1 - \rho_a - \rho_s\ .
\label{eq:5}
\end{equation}
The first expression is a quadratic equation for $\rho_a$. Its solution gives
 the average density of active sites as a function of the driving field. We can expand
 $\rho_a(h)$ for small values of $h$, i.e.
\begin{equation}
\rho_a = \rho_a^{(0)} + \chi h + O(h^2)\ ,
\label{eq:6}
\end{equation}
where $\rho_a^{(0)}$ is the zero field average density of active sites
 and $\chi$ the susceptibility, characterizing the linear response of the system
 to the external field. Substituting eq. (\ref{eq:6}) in (\ref{eq:5a}), comparing terms
 with similar order in $h$, and taking the physically admissible solutions, we obtain
\begin{equation}
\rho_a^{(0)} = 
\left\{
\begin{array}{cl}
0 &\ , \theta < \theta_c\\
\frac{1}{ug}(\theta - \theta_c) &\ , \theta > \theta_c\\
\end{array}
\right.
\label{eq:7}
\end{equation}
\begin{equation}
\chi = \frac{u}{q}|\theta - \theta_c|^{-1}\ ;
\label{eq:7a}
\end{equation}
where
\begin{eqnarray}
\theta = \frac{u(pg-1)}{pq}\ ,\ \ & \theta_c=1\ ;
\label{eq:8}
\end{eqnarray}
Moreover, just at $\theta=\theta_c$ and $h\rightarrow 0$, from eq. (\ref{eq:5a}) it 
follows that
\begin{equation}
\rho_a = \frac{1}{\sqrt{qg}}h^{1/2}\ .
\label{eq:9}
\end{equation}
Taking into account these results, from eqs. (\ref{eq:5b}) and (\ref{eq:5}) 
we obtain
\begin{equation}
\rho_s = \frac{q}{ug} + O(h,|\theta-\theta_c|)\ ,
\label{eq:9a}
\end{equation}
\begin{equation}
\rho_u = \frac{ug-q}{ug} + O(h,|\theta-\theta_c|)\ .
\label{eq:9b}
\end{equation}

Thus, for $\theta<\theta_c$, the zero field average density of active sites is zero
 while, for $\theta>\theta_c$, it is proportional to $(\theta-\theta_c)^\beta$, with
 $\beta=1$. The susceptibility diverges when $\theta\rightarrow \theta_c$,
 $\theta\neq\theta_c$, according to $|\theta-\theta_c|^{-\gamma}$, with
 $\gamma=1$. Just at $\theta_c$, the average density of active sites scales with the 
driving field as $h^{1/\delta}$, with $\delta=2$. These features are
 reminiscent of ordinary critical phenomena. $h$ is the external field, $\theta$ plays 
the role of temperature and $\rho_a$ is the order parameter. 

It is believed that conservation is a necessary condition to obtain SOC in sandpile
 models \cite{hwa,grinstein,vespignani}. The global conservation law states that the
 average input flux $hL^d$ must balance, in average, the dissipated 
flux $\rho_a\epsilon L^d$. $\epsilon$ is the dissipation rate per toppling event. It is an
 effective parameter that takes into account the dissipation, either in the bulk or at
 the boundary. Then, in the stationary state
\begin{equation}
h=\rho_a\epsilon\ .
\label{eq:12}
\end{equation}
From this equation one determines the average density of active sites as a function of
 the driving field, obtaining $\rho_a=h/\epsilon$. But, to be consistent with
 eq. (\ref{eq:6}), we must have $\rho_a^{(0)}=0$ and $\chi=1/\epsilon$. Moreover,
 using the expression for the susceptibility obtained above, eq. (\ref{eq:7a}), one
 obtains
\begin{equation}
\theta=\theta_c-\frac{u}{q}\epsilon\ . 
\label{eq:13}
\end{equation}

In systems with dissipation in the bulk $\epsilon$ is a fixed parameter and, therefore,
 the system is in a subcritical regime $\theta<\theta_c$. To obtain
 criticality ($\theta=\theta_c$) we have to fine-tune $\epsilon$ to zero. Hence, in this 
case, there is no diference with ordinary non-equilibrium critical phenomena, we
 have to fine-tune $\theta$ ($\epsilon$) to reach the critical state. On the other hand, in
 systems with dissipation at the boundary, $\epsilon$ decreases with increasing lattice
 site $L$. The number of boundary lattice sites, where dissipation may take place, 
growths slower than the total number of lattice sites. One can therefore assume 
that $\epsilon\sim L^{-\mu}$, where $\mu$ is a scaling exponent. In the 
thermodynamic limit ($L\rightarrow\infty$) $\epsilon=0$ and, from eq. (\ref{eq:13}), it
 follows that $\theta=\theta_c$. In this case, the system self-organize itself into the
 critical state. Hence, the conservation law is not a sufficient condition to obtain SOC.
 The way in which energy is dissipated plays a determinant role.

We have investigated the properties of the stationary state and the conditions to 
obtain criticality. Now we proceed to analyze the relaxation of the system after a switch
 off of the external field. If the system is close to equilibrium, $h=0$ and
 $\epsilon\approx 0$ then form eqs. (\ref{eq:2}), (\ref{eq:9b}) and (\ref{eq:13}) it follows
 that
\begin{equation}
\rho(t)\sim\exp(-p\epsilon t).
\label{eq:14}
\end{equation}
Thus the relaxation towards equilibrium is exponential, with a relaxation 
time $\tau_\epsilon\sim \epsilon^{-1}$. If the system is at the critical 
state ($\epsilon=0$) then the system takes an infinite time to reach equilibrium. This
 phenomenon is characteristic of critical systems and it is known as critical relaxation.

\subsection{Scaling laws}

The MF theory leads to an exponential relaxation, however it is well known that
 relaxation in real systems is nonexponential. The spreading of an avalanche in
 MF theory can be described by a front consisting of non-interacting particles that can 
either trigger subsequent activity or die out. The MF theory thus neglects correlations
 among the particles. Next we investigate the propagation of a deltaic perturbation, in 
space and time, though a self-organizing system close to the critical state, but using
 some scaling laws.

Let us calculate the average number of perturbed particles $\psi(t)$ due to a deltaic
 perturbation, in space and time. $\psi(t)$ is then the response function of the system.
 These particles are causally connected in space and time, thus forming an avalanche
 with average size
\begin{equation}
\langle s \rangle = \int_0^\infty dt\psi(t)\ .
\label{eq:a1}
\end{equation}
We can express the average response $\psi(t)$ as a superposition of the distribution of
 avalanche sizes, i.e.
\begin{equation}
\psi(t) = \int_0^\infty dsP(s)\psi(s,t)\ ,
\label{eq:a2}
\end{equation}
where $P(s)$ is the distribution of avalanche sizes and $\psi(s,t)$ is the reponse
 function of $s$-avalanches. In order to be consistent with eq.
 (\ref{eq:a1}) $\psi(s,t)$ must satisfy
\begin{equation}
s = \int_0^\infty dt \psi(s,t)\ .
\label{eq:a3}
\end{equation}

Now we are going to assume scaling. If $l$ is the linear dimension of
 an $s$-avalanche then its characteristic size and duration should scale
 as $l^D$ and $l^z$, respectively, where $D$ is the fractal dimension of the 
avalanches and $z$ is a dynamic scaling exponent. Therefore, the characteristic time
 scales as $s^{z/D}$ and $\psi(s,t)$ may be written as
\begin{equation}
\psi(s,t) = t^q f_1(ts^{-z/D})\ .
\label{eq:a4}
\end{equation}
where $f(x)$ is a cutoff function. The exponent $q$ is determined from the
 normalization condition in eq. (\ref{eq:a3}), obtaining $q=D/z-1$. Moreover, 
$\psi(s,t)$ is independent of $s$ for $ts^{-z/D}\ll 1$ then $f(x)\sim\text{cte}$ for
 $x\ll 1$. On the other hand, $\psi(s,t)\ll 1$ for $ts^{-z/D}\gg 1$ and, therefore,
 $f(x)\ll 1$ for $x\gg 1$. 

The distribution of avalanche sizes near the SOC state satisfy the scaling relation
\begin{equation}
P(s)=s^{-\tau}f_2(s\epsilon^{1/\sigma})\ ,
\label{eq:a5}
\end{equation}
where $\tau$ and $\sigma$ are scaling exponents and $f_2(x)$ is a cutoff function with
 similar asymptotic behavior as $f_1(x)$. $\epsilon$ is the control parameter which 
determine how close is the system to the SOC state, i.e. the dissipation rate in the previous
 MF approach.

Substituting eqs. (\ref{eq:a4}) and (\ref{eq:a5}) in eq. (\ref{eq:a2}) one obtains
\begin{equation}
\psi(t)=t^{\kappa-1}f_3(t\epsilon^{1/\Delta})\ ,
\label{eq:a6}
\end{equation}
where
\begin{equation}
\Delta=\frac{D}{z}\sigma\ ,\ \ \ \ \kappa=\frac{D(2-\tau)}{z}\ ,
\label{eq:a7}
\end{equation}
and $f_3(x)$ is a cutoff function.

Finally, the normalized relaxation function $\phi(t)$ defined through the expression
\begin{equation}
\phi(t) = \frac{\int_t^\infty dt^\prime \psi(t^\prime)}  
{\int_0^\infty dt^\prime \psi(t^\prime)}\ ,
\label{eq:a8}
\end{equation}
may be approximated, for $t\epsilon^{1/\Delta}\ll 1$, by
\begin{equation}
\phi(t)\sim\exp\Big[-\Big(\frac{t}{\tau_\epsilon}\Big)^\kappa\Big]\ ,
\label{eq:a9}
\end{equation}
where $\tau_\epsilon\sim\epsilon^{-1/\Delta}$.
Thus, under the scaling laws assumed above the relaxation function follows the
 Kohlrausch stretched exponential. 

The stretched exponent $\kappa$ is defined in eq. (\ref{eq:a7}) through the scaling
 exponents $\tau$, $D$ and $z$. However, using some scaling relations we can obtain
 a  simpler expression. In systems at a SOC state the correlation length scales as
 the system size $\xi\sim L$ and $\xi\sim\epsilon^{-\nu}$, where $\nu$ is a scaling
 exponent. It is also possible to show that in homogeneous SOC systems
 $\langle s\rangle\sim L^2$. Moreover, from eq. (\ref{eq:a5}) it follows that 
$\langle s\rangle \sim \epsilon^{-(2-\tau)/\sigma}$. Hence,
\begin{equation}
\langle s\rangle \sim L^2 \sim \xi^2 \sim \epsilon^{-2\nu} \sim \epsilon^{-(2-\tau)/\sigma},
\label{eq:a10}
\end{equation}
and therefore
\begin{equation}
2-\tau = 2\sigma\nu.
\label{eq:a11}
\end{equation}
On the other hand, the cutoff avalanche size is given by
\begin{equation}
s_c \sim \xi^D \sim \epsilon^{-D\nu} \sim \epsilon^{-1/\sigma},
\label{eq:a12}
\end{equation}
and therefore
\begin{equation}
D\nu\sigma = 1.
\label{eq:a13}
\end{equation}
Then, using (\ref{eq:a7}), (\ref{eq:a11}) and
 (\ref{eq:a13}), one obtains the scaling relations
\begin{equation}
\Delta=\frac{1}{z\nu}\ ,\ \ \ \ \kappa=\frac{2}{z}\ .
\label{eq:a14}
\end{equation}

The stretched exponent $\kappa$ only depends on the dynamic scaling exponent
 $z$ (the duration of an avalanche scales with its linear dimension as $l^z$). In the MF
 approach $z=2$ in order to recover the exponential relaxation obtained in the
 preceding subsection. However, if one considers correlations among the different
 branches forming an avalanche then one expects that the duration of the 
avalanche grows faster with its linear dimension and, therefore, $z>2$.

\section{Discussion}

We have obtained that the normalized relaxation function follows an stretched
 exponential decay, with a stretched exponent given by the second scaling relation in
 eq. (\ref{eq:a14}). However, this functional dependence is limited to the time scale
 $1\ll t\ll \tau_\epsilon\sim\epsilon^{1/\Delta}$. In the SOC
 state $\epsilon=0$ and, therefore, the stretched exponential decay will be observed 
for infinitely long times.

This temporal behavior leads to the frequency spectrum
\begin{equation}
S(f) = \int_0^\infty dt \cos(2\pi ft) \phi(t) \sim f^{-1-\kappa},
\label{eq:b1}
\end{equation}
in the frequency range $\epsilon^{-1/\Delta}\ll f\ll 1$. We thus found that the
 avalanche dynamics leads to a ''$1/f$'' noise in the intermediate frequency window.
 This result seems to be similar to that reported by Bak, Tang and Wiesenfeld in their
 pioneering work \cite{bak}, however this is not the case. 
Bak, Tang and Wiesenfeld obtained a ''$1/f$'' noise assuming a power law distribution 
$P(T)\sim T^{-\alpha}$ of avalanche durations, which is equivalent to assume 
a power law distribution of avalanches sizes, but they also assume
 that each avalanche relaxes exponentially, i.e.
\begin{equation}
\phi(t)=\int_0^\infty P(T)\exp(t/T)\sim t^{-(\alpha-1)},
\label{eq:b2}
\end{equation}
thus obtaining $S(f) \sim f^{-2+\alpha}$. The exponent $\alpha$ may be expressed in
 terms of of the scaling exponents introduced here.
 Since $T\sim s^{z/D}$ and $P(s)ds=P(T)dT$ then
\begin{equation}
\alpha = 1+\frac{D}{z}(\tau-1) = 1 + \frac{D-2}{z},
\label{eq:b4}
\end{equation}
where the second equality is obtained using the scaling relations in equations
 (\ref{eq:a11}) and (\ref{eq:a13}).

In both cases the power spectrum is of the form 
$1/f^\beta$, but $\beta=1+\kappa$ is larger than one in our model, 
while $\beta=1-(D-2)/z$ is smaller than one in the Bak, Tang and Wiesenfeld
 approach. Moreover, the functional dependence of the normalized relaxation function 
is different, we obtain a stretched exponential while they obtain a power law decay. 
This discrepancy is a consequence of the assumption made by Bak, Tang 
and Wiesenfeld \cite{bak} that the avalanches relax exponentially, with a relaxation
 time equal to its duration. Such an assumption does not take into account the intrinsic
 dynamic nature of an avalanche, which may lead by itself to a nonexponential 
relaxation. 

The normalized relaxation function associated with an avalanche of duration $T$,
 or equivalently of size $s$ ($T\sim s^{z/D}$), can be obtained from eqs. (\ref{eq:a4}) 
and (\ref{eq:a8}), resulting
\begin{equation}
\phi(T,t) = \frac{\int_{(t/T)}^\infty dx x^{1-z/D} f_1(x)}{\int_0^\infty dx x^{1-z/D} f_1(x)}.
\label{eq:b5}
\end{equation}
Hence, the characteristic time is the avalanche duration, but the normalized relaxation
 function is not necessarily exponential. It is well known that cooperative dynamics l
eads, without assuming any disorder, to a nonexponential relaxation dynamics.

\section{Conclusions}
\label{sec:conclusions}
We have proposed the avalanche dynamics as the origin of the ubiquitous 
nonexponential relaxation observed in complex systems. Introducing some scaling 
laws and relations we have obtained that the normalized relaxation function follows 
an stretched exponential decay with an stretched exponent
 $\kappa=2/z$, where $z$ is the dynamic scaling exponent. Moreover, in the MF
 approach $\kappa=1$ and the relaxation is exponential.

The stretched exponential decay leads to the ''$1/f$'' noise spectrum
 $S(f)\sim f^{-1-\kappa}$. This result is more general than the one reported by
 Bak, Tang, and Wiesenfeld \cite{bak}, since it takes into account the intrinsic 
dynamic nature of the avalaches.

\section*{Acknowledgements}

This work was partially supported by the {\em Alma Mater} prize,
given by The University of Havana.
One of us (O.S) wants to acknowledge the kind hospitaity of UNED. Financial support
 by the Ministerio de Educación y Ciencias  is gratefully acknowledged

\end{multicols}

\end{document}